\documentclass[aps,twocolumn,a4paper]{revtex4}
\usepackage{tensor}
\usepackage{graphicx}
\usepackage{amsmath}
\usepackage{amssymb}
\usepackage{enumerate}
\usepackage{subfigure}
\usepackage{tabularx}
\usepackage[colorlinks=true, pdfstartview=FitV, linkcolor=blue, citecolor=red, urlcolor=black, breaklinks=true]{hyperref}
\usepackage{amsthm}
\usepackage{epstopdf}
\newcommand{\be}{\begin{equation}}
\newcommand{\ee}{\end{equation}}
\newcommand{\ben}{\begin{eqnarray}}
\newcommand{\een}{\end{eqnarray}}
\newcommand{\bes}{\begin{subequations}}
\newcommand{\ees}{\end{subequations}}
\def\bal#1\eal{\begin{align}#1\end{align}}

\newcommand{\sech}{{\rm sech}}

\newcommand{\Lag}{\mathcal{L}} 
 
\newcommand{\pu}{\mathrm{\partial_{\mu}}}

\newcommand{\Vfc}{V(\phi, \chi)}
\newcommand{\f}{{f}(\chi)}
\begin{document}
\title{Geometrically Constrained Kinklike Configurations}
\author{D. Bazeia}\affiliation{Departamento de F\'\i sica, Universidade Federal da Para\'\i ba, 58051-970 Jo\~ao Pessoa, PB, Brazil}
\author{M. A. Liao}\affiliation{Departamento de F\'\i sica, Universidade Federal da Para\'\i ba, 58051-970 Jo\~ao Pessoa, PB, Brazil}
\author{M. A. Marques}\affiliation{Departamento de F\'\i sica, Universidade Federal da Para\'\i ba, 58051-970 Jo\~ao Pessoa, PB, Brazil}
\begin{abstract}
In this work we study kinklike structures, which are localized solutions that appear in models described by real scalar fields. The model to be considered is characterized by two real scalar fields and includes a function of one of the two fields that modifies the kinematics associated to the other field. The investigation brings to light a first order framework that minimizes the energy of the solutions by introducing an auxiliary function that directly contributes to describe the system. We explore an interesting route, in which one field acts independently, entrapping the other field, inducing important modifications in the profile of the localized structure. The procedure may make the solution to spring up as a kinklike configuration with internal structure, engendering the important feature that also appears directly connected with issues of current interest at the nanometric scale, in particular in the electronic transport in molecules in the presence of vibrational degrees of freedom. 
\end{abstract}
\maketitle

Spontaneous symmetry breaking is ubiquitous in physics. It is deeply connected with phase transitions and the generation of localized structures in nature. These structures engender finite energy, may attain topological behavior and usually appear in one, two and three spatial dimensions. In high energy physics they are called kinks, vortices and magnetic monopoles, respectively; see, e.g., \cite{B1,B2,B3} and references therein. In the present study we concentrate on kinks, which in the simplest form are generated under the presence of a single real scalar field in $(1,1)$ spacetime dimensions. The model usually engenders a standard kinetic term and a potential, which controls the nonlinearity of the physical system and develop spontaneous symmetry breaking.

Since the pioneer work of Finkelstein \cite{P}, who coined the word kinks, these localized structures have been studied in a diversity of contexts in physics, for instance, in high energy physics \cite{K0,K1,K2,K3,K4,K5}, in condensed matter \cite{KCM1,KCM2,KCM2a,KCM3,KCM4} and in other areas \cite{O1,O2,O3,O4,O5} of nonlinear science. In high energy physics, in \cite{K0} the authors described a procedure to construct new models capable of supporting kinklike solutions. Also, in \cite{K1} kinks have been studied in $(D,1)$ spacetime dimensions, that is, in arbitrary $D$ spatial dimensions, in a way circumventing the Derrick-Hobbard scaling theorem \cite{D,H}, which states that a scalar field with standard kinematics cannot provide kinklike configurations, unless we work in $(1,1)$ spacetime dimensions. There we have also described the presence of solutions with the profile of a two-kink configuration. Kinks also appeared in \cite{K2}, in the study of the creation of solitons from particles, in which the scattering of wave pulses creates kink-antikink pairs; in \cite{K3}, which describes the presence of a new kink of a massive nonlinear sigma model with $S^2$ sphere as the target manifold; in \cite{K4}, where complex (twisted) kinks are shown to appear as exact self-consistent solutions in Bogoliubov–de Gennes and chiral Gross-Neveu systems; and in \cite{K5}, in a investigation that describes the decay of cosmic string loops in the Abelian-Higgs model as primarily due to kink collisions. 

In condensed matter, in \cite{KCM1} the authors found two-kink structures experimentally, in the micrometer-sized $\rm {Fe}_{20} {\rm Ni}_{80}$ magnetic material under the presence of constrained geometries; in \cite{KCM2} it was shown that a kinklike configuration in
$\rm {Fe}_{30} {\rm Ni}_{70}$ magnetic nanowires may change polarity under the presence of an electric current; in \cite{KCM2a}, which reported on the longitudinal spin Seebeck effect, measured in a system composed of a ferrimagnetic insulator $\rm{Y}_3\rm{Fe}_5\rm{O}_{12}$ slab and a
$\rm{Pt}$ film by means of the inverse spin-Hall effect, with the voltage acquiring the profile of a two-kink configuration; in \cite{KCM3}, where the investigation focused on the propagation of a domain wall under the effect of a magnetic field, to lead to the formation of kinks, which can behave like sine-Gordon solitons in thin films of materials such as yttrium iron garnets; and in \cite{KCM4}, where the authors established an analogy between the excitations of a buckled graphene nanoribbon and kinks of the well-known $\phi^4$ model, the prototype of the Higgs model.

In other areas of nonlinear science, in \cite{O1,O2,O3,O4}, for instance, the electronic transport in molecules may generate a current with the two-kink behavior, which can be further modified to have a novel profile, which appears when the investigation includes vibrational degrees of freedom of the molecule. Also, in \cite{O5} the authors observed that the base pairing on the conformations of RNA undergoes a continuous transition from swollen coil to globule with a profile in the shape of a kink.

The two-kink solution that appeared before in \cite{KCM1} is directly related to the presence of the geometric constriction there introduced.  A particular geometric junction was also used in \cite{sky1} to create skyrmions from domain-walls or kinks. Another situation where a geometric constriction is of key importance, was explored in \cite{sky2}, and there, the authors demonstrated experimentally the current-driven transformation of kinklike configurations into magnetic skyrmions in a magnetic strip, an issue of direct interest to skyrmion-based spintronics. These results suggest that the study of kinks at the nanometric scale needs further attention, because the presence of geometrical constrictions may induce internal modifications in the profile of the localized structure, contributing to change the physical properties of the system under investigation. Similar effect appeared in the voltage in the longitudinal spin Seebeck effect \cite{KCM2a}, and in the electric current through molecules \cite{O1,O2,O3}, which also develop the two-kink profile, or the profile shown in \cite{O2,O3} when one adds vibrational degrees of freedom; see also \cite{O6}. 

The above facts motivated us to concentrate on the study of kinks, searching for the presence of kinklike configurations with internal structure, taking into account new effects, capable of simulating geometrical modifications that may appear at the nanometric scale. Since in its simplest form the kink requires a real scalar field and the presence of the $Z_2$ symmetry, we modify the system and enlarge the symmetry to the $Z_2\times Z_2$ case, with the addition of another real scalar field to control the extra symmetry. In this sense, we work with the action of two real scalar fields, $\phi$ and $\chi$, in $(1,1)$ spacetime dimensions with metric tensor $\eta_{\mu\nu}={\rm diag}(1,-1)$ and natural units such that $\hbar=c=1$. For simplicity, we also consider all the fields, parameters and space and time coordinates dimensionless, to help us focus on the main aspects of the problem; the addition of dimensions can be implemented standardly. The action follows the usual definition, and the Lagrange density is given by
	\begin{equation}\label{lagr}
		\Lag = \frac{1}{2}f(\chi)\pu \phi \partial^{\mu} \phi + \frac{1}{2}\pu \chi \partial^{\mu} \chi  - V(\phi,\chi).
	\end{equation}
The function $f(\chi)$ is nonnegative, and modifies the kinematics of the field $\phi$. The scalar fields are, in principle, coupled through the potential $\Vfc$ and the function $f(\chi)$. One may use $\phi\to\tilde\phi+\delta \phi$ and $\chi\to\tilde\chi+\delta \chi$ to vary the action associated to the above model to get the equations of motion. Also, invariance of the action under translations in the spacetime leads to the energy-momentum tensor $T_{\mu\nu}$. We implement a detailed investigation focusing on the search of kinks considering static configurations; in this case, the equations of motion are
	\begin{subequations}\label{eoms}
		\begin{align}
			\label{est1} &\frac{d}{dx}\left( f(\chi)\frac{d\phi}{dx}\right)=V_{\phi}{(\phi,\chi)}, \\ 
			\label{est2}&\frac{d^2\chi}{dx^2} - \frac{1}{2}\frac{df(\chi)}{d\chi}\left(\frac{d\phi}{dx} \right)^ 2 = V_{\chi}{(\phi,\chi)},
		\end{align}
	\end{subequations}
where $V_{\phi}= \partial V/\partial \phi$, $V_{\chi}= \partial V/\partial \chi$ and $f_{\chi}= d f/d \chi$.
Asymptotically, the two field solutions have to obey $\phi(x\to\pm\infty)\to v_\pm$ and $\chi(x\to\pm\infty)\to w_\pm,$
where $v_\pm$ and $w_\pm$ are parameters related to the form of the potential, usually identifying its minima. In particular, when $\chi$ becomes $w_+$ or $w_-$, the potential only depends on $\phi$ and if we impose that $f(w_\pm)=1$, the model turns out to describe a single real scalar field, with standard kinematics; see below for other details on this.  In the presence of static configurations, the non-vanishing components of the energy-momentum tensor are $T_{00}$ and $T_{11}$, the energy density and the stress associated to the static solutions. 
The energy density has the form
\begin{equation}
 \rho\equiv T_{00}=\frac{1}{2}\f\left(\frac{d\phi}{dx}\right)^2 +\frac12 \left(\frac{d\chi}{dx}\right)^2 + \Vfc. \label{T00} \\
\end{equation}
We work with Ref.~\cite{B}, using the Bogomol'nyi procedure to investigate the conditions under which the model gives rise to a first order framework. We go further on this and introduce an auxiliary function $W(\phi,\chi)$ such that \eqref{T00} can be rewritten in the form
\begin{equation}\label{bog}
	\begin{aligned}
\rho &=\frac{f(\chi)}{2}\left(\frac{d\phi}{dx} \mp \frac{W_{\phi}}{f(\chi)}\right)^2 + \frac{1}{2}\left(\frac{d\chi}{dx} \mp W_{\chi}\right)^2 \\
 &\hspace{4mm}+ V-\left(\frac{1}{2}\frac{W_{\phi}^2}{\f}+\frac{1}{2}W_{\chi}^2\right) \pm\frac{dW}{dx},
\end{aligned}
\end{equation}
where $W_\phi=\partial W/\partial\phi$ and $W_\chi=\partial W/\partial\chi$. We then write the potential in the form
	\begin{equation}\label{potgen}
		\Vfc=\frac{W_{\phi}^2}{2f(\chi)} + \frac{W_{\chi}^2}{2},
	\end{equation}
and this leads to the conclusion that the energy is now bounded by $E \geq E_B \equiv \left|W(v_+,w_+)-W(v_-,w_-)\right|$. 
Moreover, the energy is minimized to $E_B$ if the scalar fields solve the first order equations
	\begin{equation}\label{fo}
			\frac{d\phi}{dx}=\pm \frac{W_{\phi}(\phi, \chi)}{f(\chi)},\;\;\;\;\;
			\frac{d\chi}{dx}=\pm W_{\chi}(\phi,\chi).
	\end{equation}
An interesting fact is that, even though the function $f(\chi)$ appear in the equations of motion \eqref{eoms} and in the above first order equations, it does not contribute to the energy, which only depends on the function $W$ and the asymptotic values of the field configurations. One can also check that the solutions of the above first order equations \eqref{fo} satisfy the second order equations of motion \eqref{eoms}.

We observe that the system of first order equations becomes particularly interesting when $W(\phi,\chi)$ is written as the sum of two terms, $W(\phi,\chi)=W_1(\phi)+ W_2(\chi)$, such that $W_{\phi\chi}=W_{\chi\phi}=0$. In this case the second of the first order equations \eqref{fo}
does not depend on $\phi$, so the $\chi$ field can be solved independently, to feed the behavior of the other field, $\phi$. In this work we follow this strategy, that is, we use the field $\chi$ to act as the agent to change the behavior of the other field. In this sense, since a kink is a topological structure which is localized around its center, we can make the $\chi$ field to generate a kink, and use it to modify the other field configuration, so the model can be used to entrap the field $\phi$, working to unveil effects that can possibly appear under the presence of geometric constrictions. As we will show below, our proposal leads to the presence of kinklike solutions with internal structures that are novel field configurations with no precendent in the related literature. In fact, we innovate in the sense that the new structures springs up in the presence of constrained geometries, so they are of current interest, directly connected to the study of localized structures at the nanometric scale. 

The existence of the first order equations \eqref{fo} allows us to distinguish two contributions to the energy density, given by
$\rho=\rho_1(\phi(x),\chi(x)) + \rho_2(\chi(x)),$ where
\begin{equation}\label{foeq}
\rho_1=f(\chi)\left(\frac{d\phi}{dx}\right)^2,\;\;\;\;\;
\rho_2=\left(\frac{d\chi}{dx}\right)^2.
		\end{equation}
Since the solutions of the first order equations have energy minimized to $E_B$, they are stable against small fluctuations of the scalar fields \cite{B}. We remark that the stress component $T_{11}$ vanishes for field configurations that solve the first order equations \eqref{fo}, so the solutions are also stable under rescaling \cite{D,H,genkink,genkink2}.

The above procedure introduces an interesting approach to deal with kinks, which can be used to describe distinct properties of the solutions. We further explore this possibility considering some distinct examples which capture the essence of the proposal. We first suppose that $f(\chi)=1/\chi^2$ and take the function
\begin{equation}\label{wchi4}
W(\phi,\chi) =\phi - \frac13\phi^3 + \alpha \chi - \frac13 {\alpha}\chi^3,
\end{equation}
where $\alpha$ is a non-negative real parameter. Although the above $W(\phi,\chi)$ allows the presence of four distinct parameters, here one only uses $\alpha$, which is enough to expose the main feature of the model, helping us to control the profile of the solutions. The potential in Eq.~\eqref{potgen} becomes
	\begin{equation}\label{pot1}
	V(\phi,\chi)=\frac12{\chi^2}(1-\phi^2)^2+\frac12\alpha^2(1-\chi^2)^2.
	\end{equation}
It is nonnegative, so the absolute minima are at the values $\pm 1$, which identify both $v_\pm$ and $w_\pm$, needed to describe the boundary conditions for the solutions. We note that $f(\pm1)=1$, as expected. So, for $\chi=\pm1$ the model describes the standard $\phi^4$ model, the prototype of the Higgs model, supporting the standard kinklike solution, with the hyperbolic tangent form. 

		\begin{figure}[h!]
		\centering
		\includegraphics[width=4.2 cm]{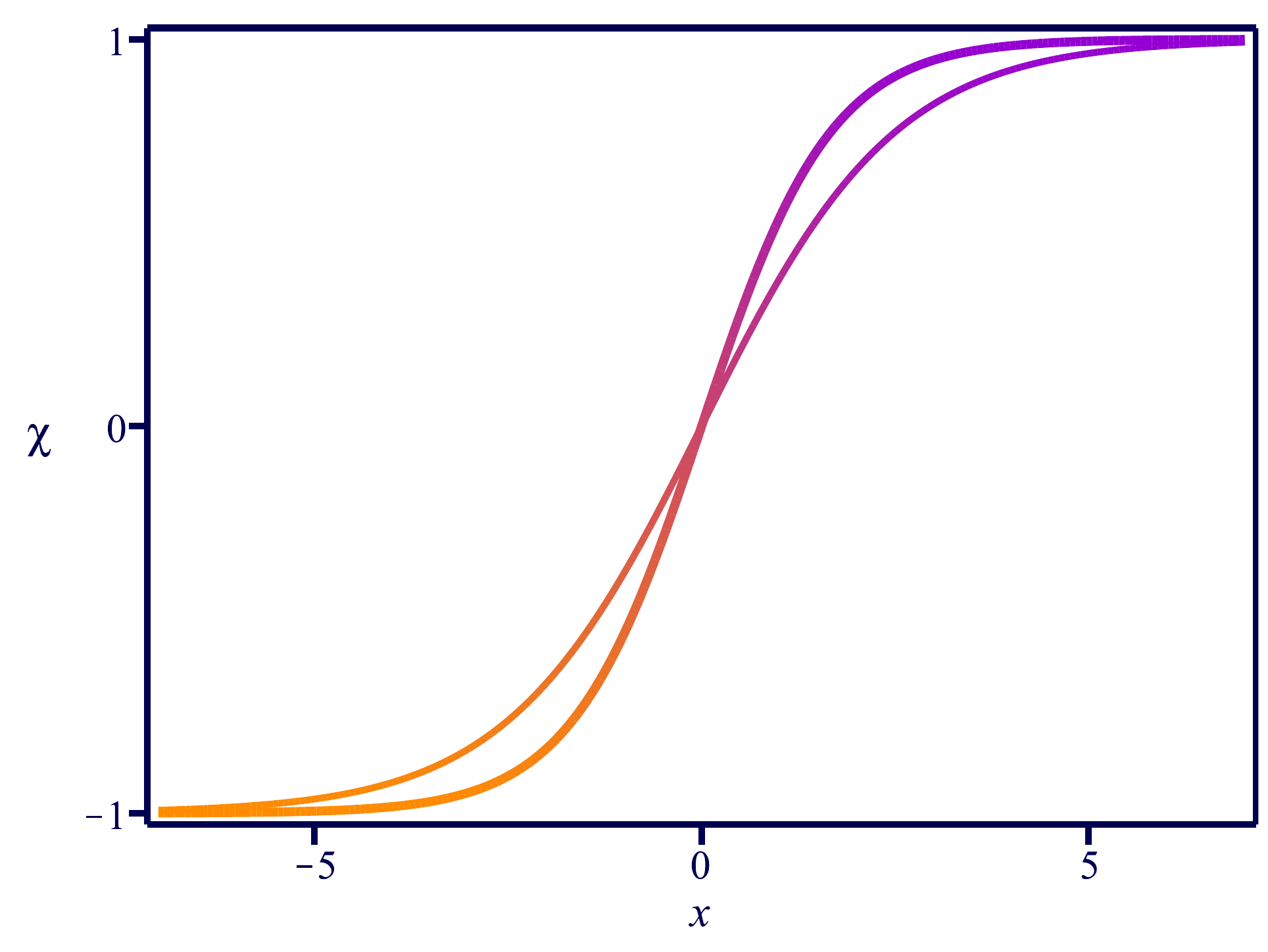}
		\includegraphics[width=4.2 cm]{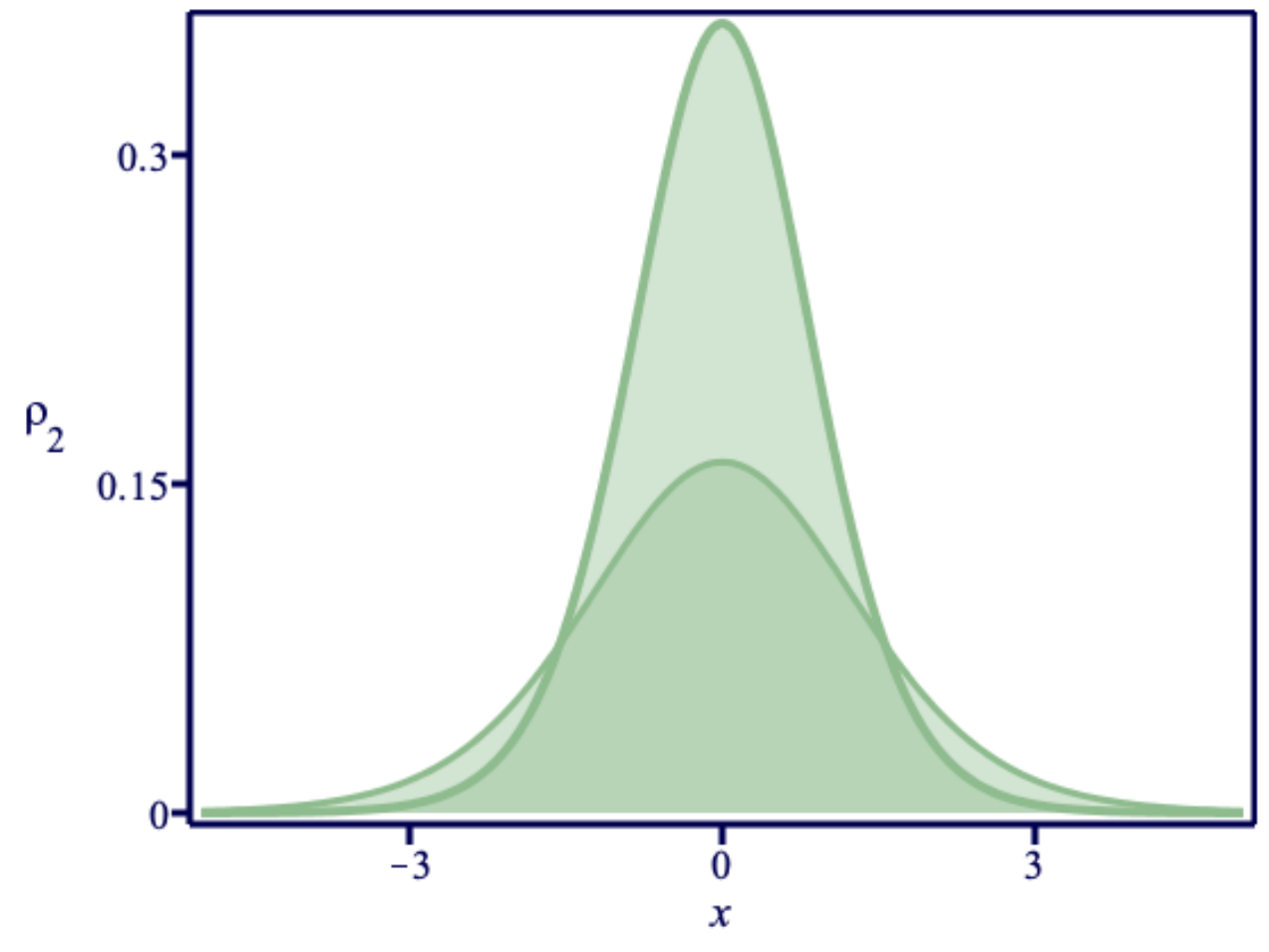}
		\caption{The solution $\chi(x)$ (left) and the energy density $\rho_2(x)$ (right), depicted for
		$\alpha = 0.4$ and $0.6$. The thickness of the lines increases with $\alpha$.}
		\label{fig1}
		\end{figure}

In the presence of static solutions, the above potential minimizes the energy to the value $E_B=4(1+\alpha)/3$, and the first order equations become
	\begin{equation}
		\label{firstorder1}\frac{d\phi}{dx}=\pm \chi^2 (1- \phi^2),\;\;\;\;\;\frac{d\chi}{dx}=\pm \alpha (1 - \chi^2).	\end{equation}
These equations support kinks and anti-kinks but, since they are related by a reflection in the $x$ coordinate, we only consider the case of kinks. As one knows, the first order equation for $\chi$ describes the standard $\chi^4$ model, with solution $
\chi(x)= \tanh(\alpha x).$ Near the origin, $\chi(x)\propto \alpha x$, so $\alpha$ contributes to modify the behavior of the solution near its center located at the origin. The contribution of this field to the energy density $\rho_2$ appears in Eq.~\eqref{foeq}, giving $\rho_2(x)=\alpha^2\sech^4(\alpha x)$. After integration we get $E_2 = 4\alpha/3$. The profile of this kink and the corresponding energy density $\rho_2(x)$ can be seen in Fig.~\ref{fig1}. One notices that as $\alpha$ increases, the solution becomes more localized around its center, located at the origin.
So, the $\chi$ kink may work to entrap the other field, $\phi$, and the parameter $\alpha$ may act to make the entrapment more or less significant. 

\begin{figure}[h]
\centering
\includegraphics[width=4.2cm]{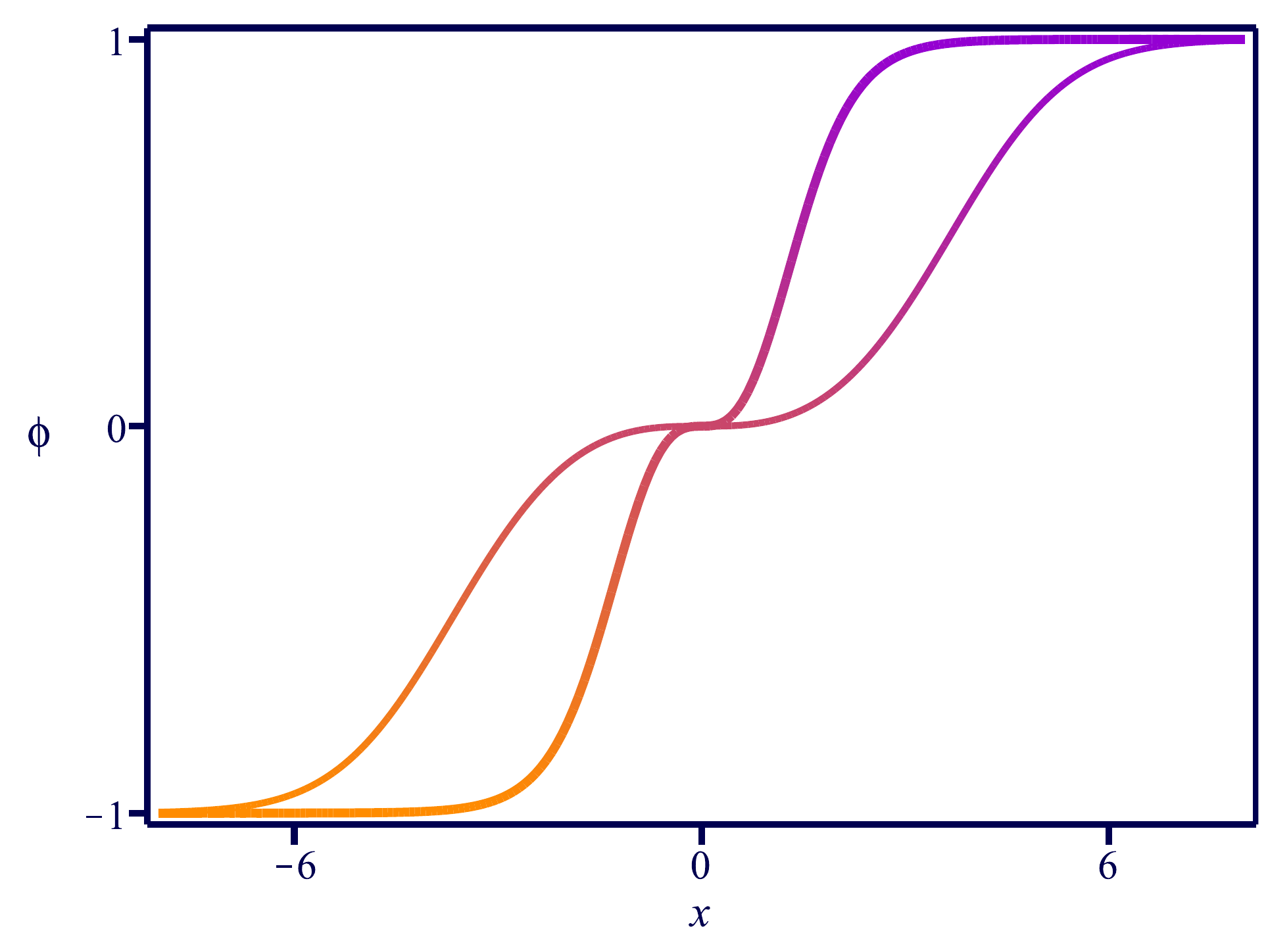}
\includegraphics[width=4.2cm]{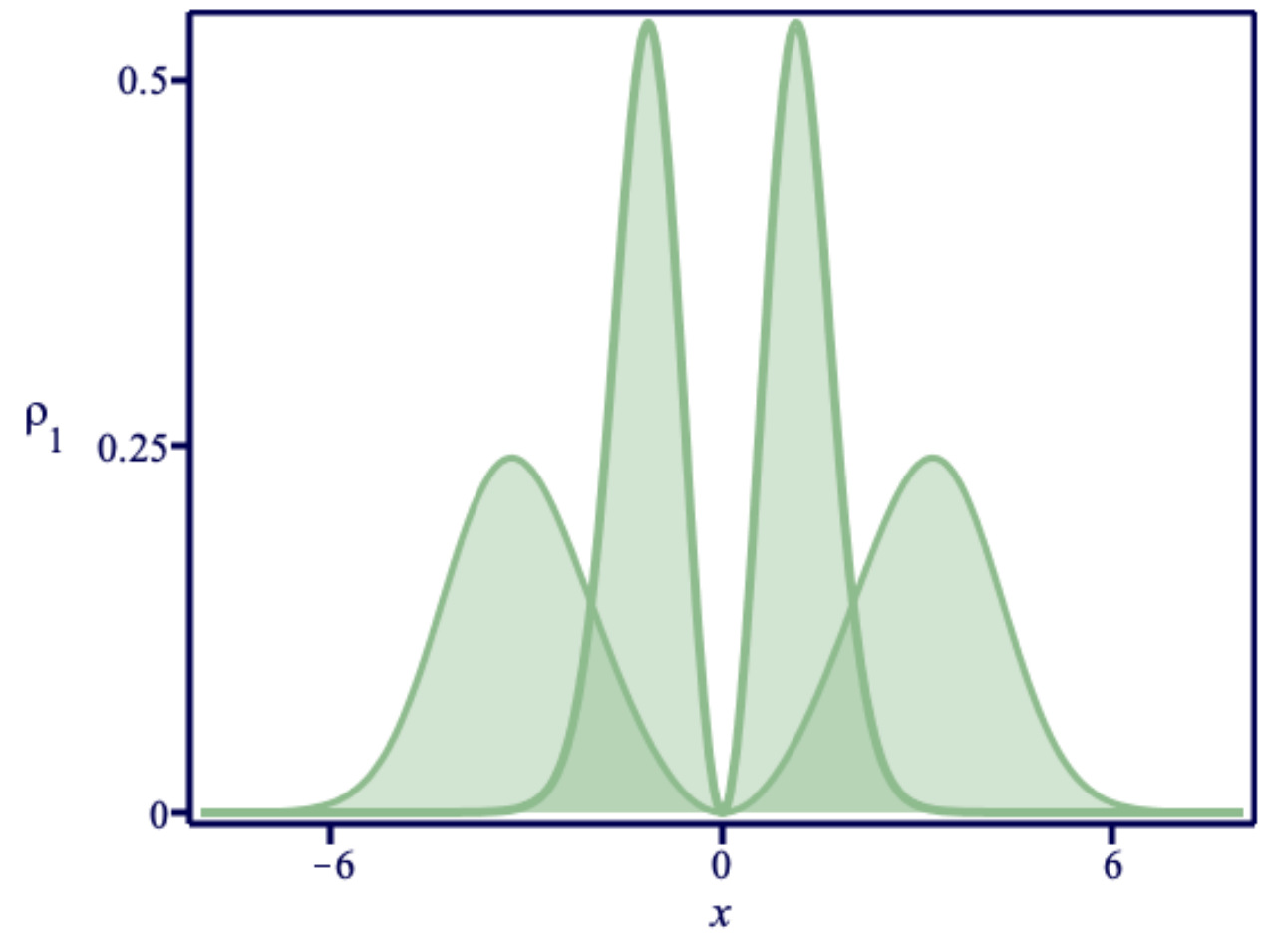}
\caption{The solution $\phi(x)$ (left) and the energy density $\rho_1(x)$ (right),
for $\alpha =0.2$ and $1$. The thickness of the lines increases with $\alpha$.}
\label{fig2}
\end{figure}

After substituting the solution $\chi(x)$ in \eqref{firstorder1} with the upper sign, we obtain the first order equation for $\phi$ 
	\begin{equation}\label{fophi}
	\frac{d\phi}{dx}= \tanh^2(\alpha x) (1 - \phi^2),
	\end{equation}
which admits the solution
	\begin{equation}\label{phi1}
		\phi =\tanh(\xi(x)),
	\end{equation}
where $\xi(x)\equiv x - \tanh(\alpha x)/\alpha$. This solution is shown in Fig.~\ref{fig2}. Interestingly, it has the two-kink profile, similar to the case found experimentally in \cite{KCM1} or in \cite{KCM2a}. Near the origin, the solution behaves as $\phi(x)\propto \alpha^2 x^3$, indicating the presence of a plateau at the center of the solution, as it happens for the two-kink configuration. The parameter $\alpha$ controls the behavior of the solution around the origin, making the plateau narrower as it is increased; see Fig. \ref{fig2}.
The contribution $\rho_1(x)$ to the energy density is given by Eq.~\eqref{foeq}, which becomes
\begin{equation}\label{rphi1}
		\rho_1(x)=\sech^4(\xi(x))\tanh^2(\alpha x).
	\end{equation}
It can be integrated to give $E_1=4/3$. The sum $E_1+E_2$ gives the total energy, $E=4(1+\alpha)/3$, which matches with the value $E_B$ obtained below Eq. \eqref{pot1}, as expected. The energy density $\rho_1(x)$ is also depicted in Fig.~\ref{fig2} for some values of $\alpha$. 
\begin{figure}[t!]
\centering{
\includegraphics[width=4.3cm]{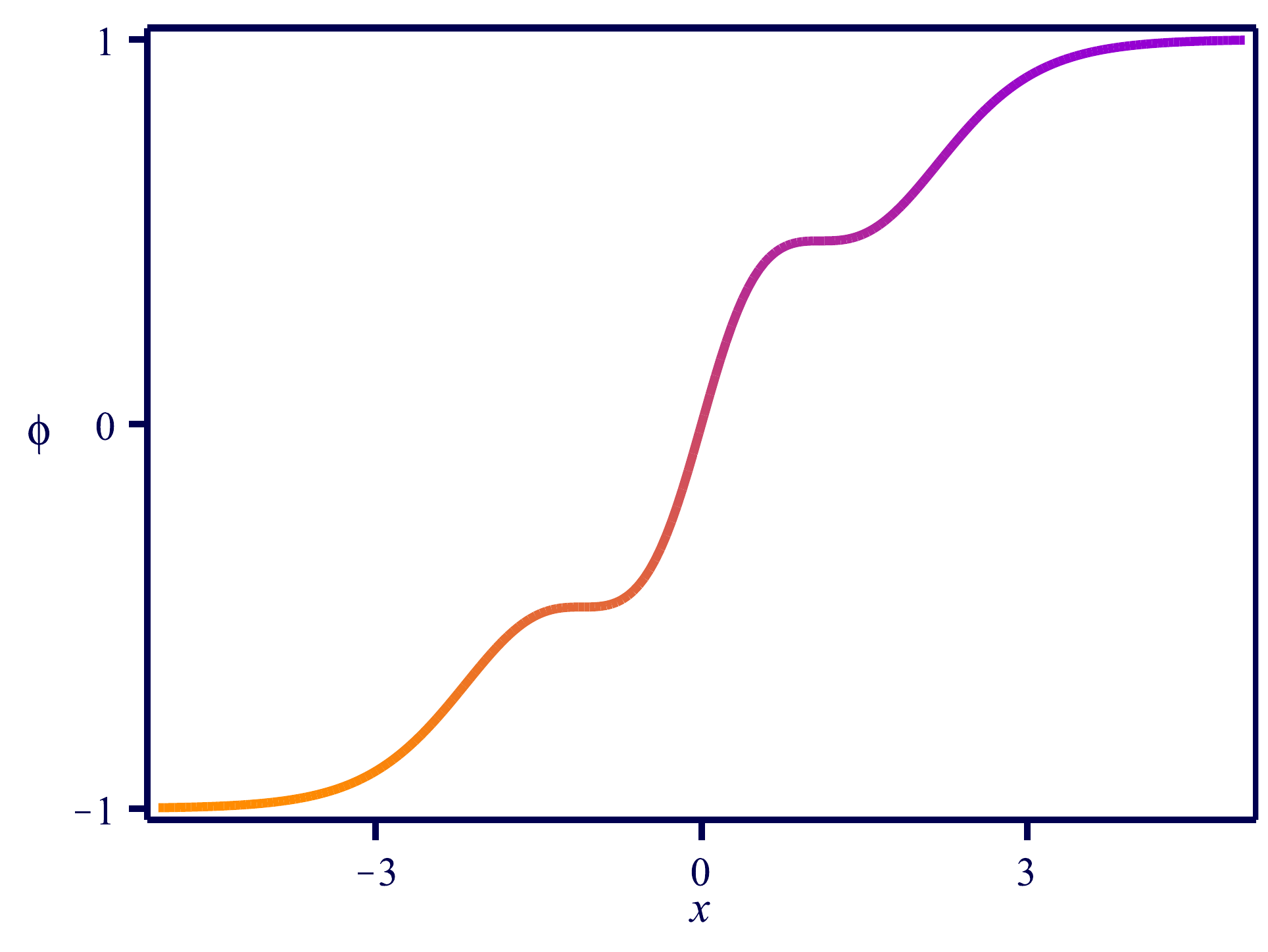}\includegraphics[width=4.3cm]{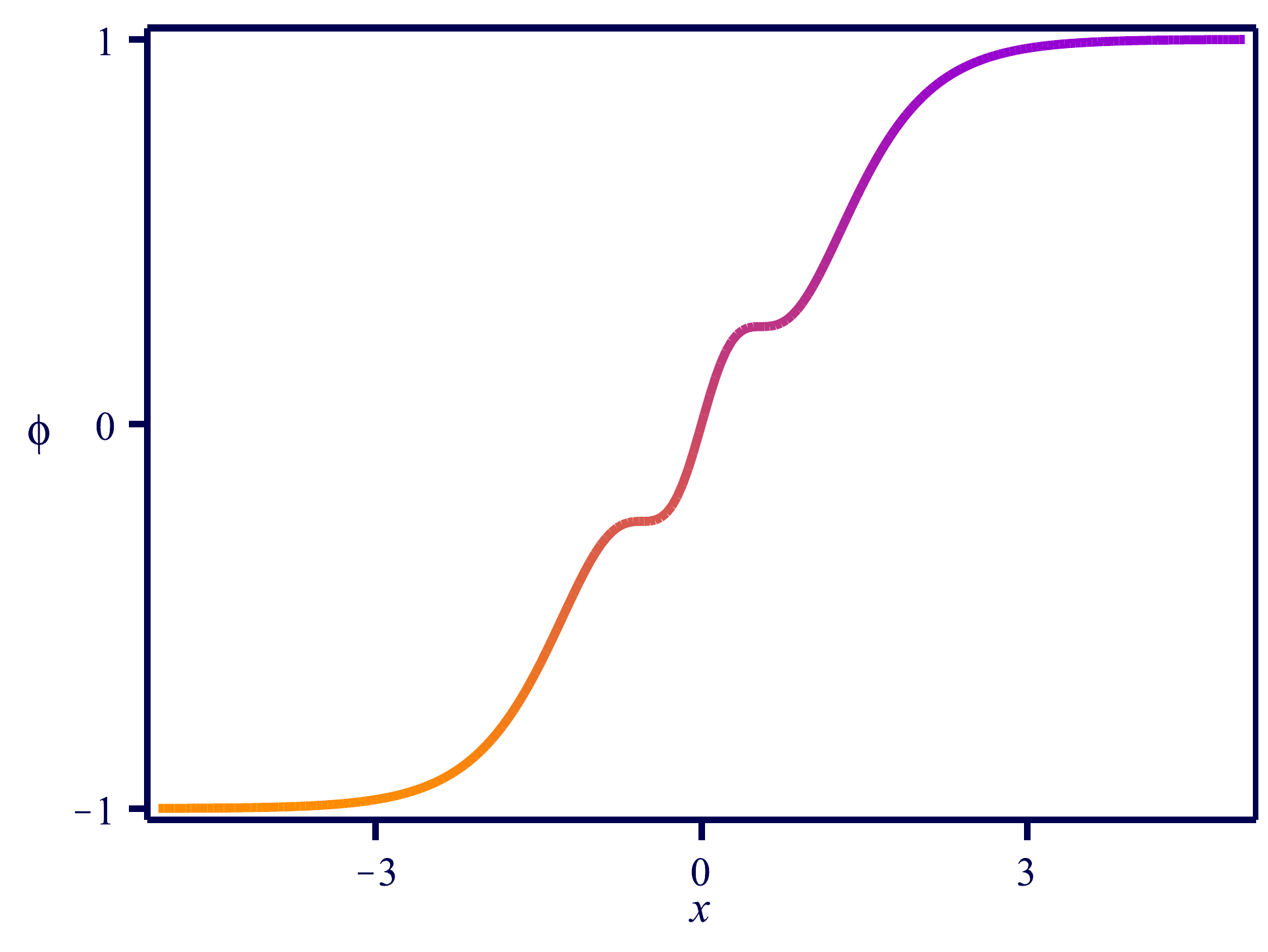}
\includegraphics[width=4.3cm]{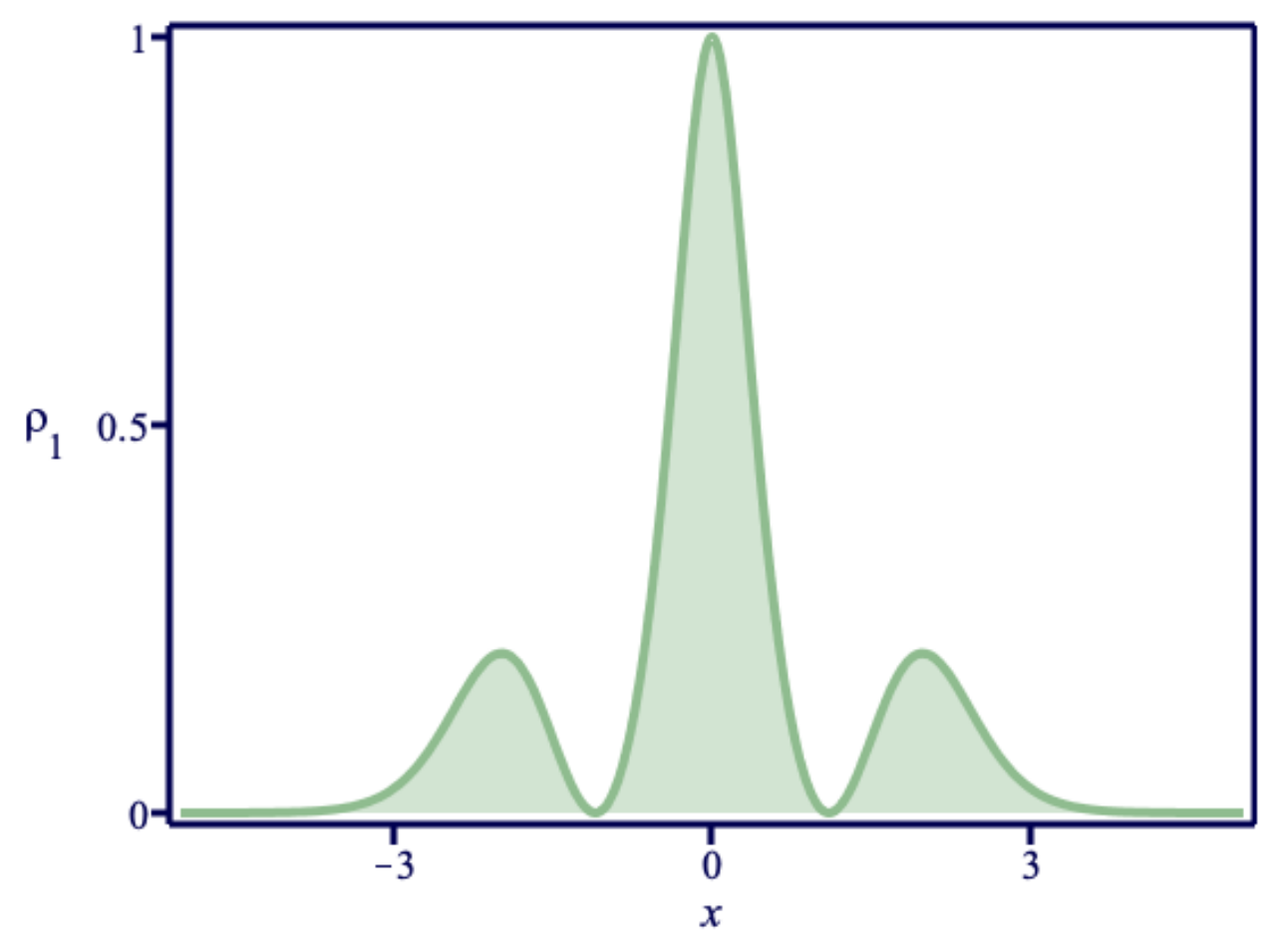}\includegraphics[width=4.3cm]{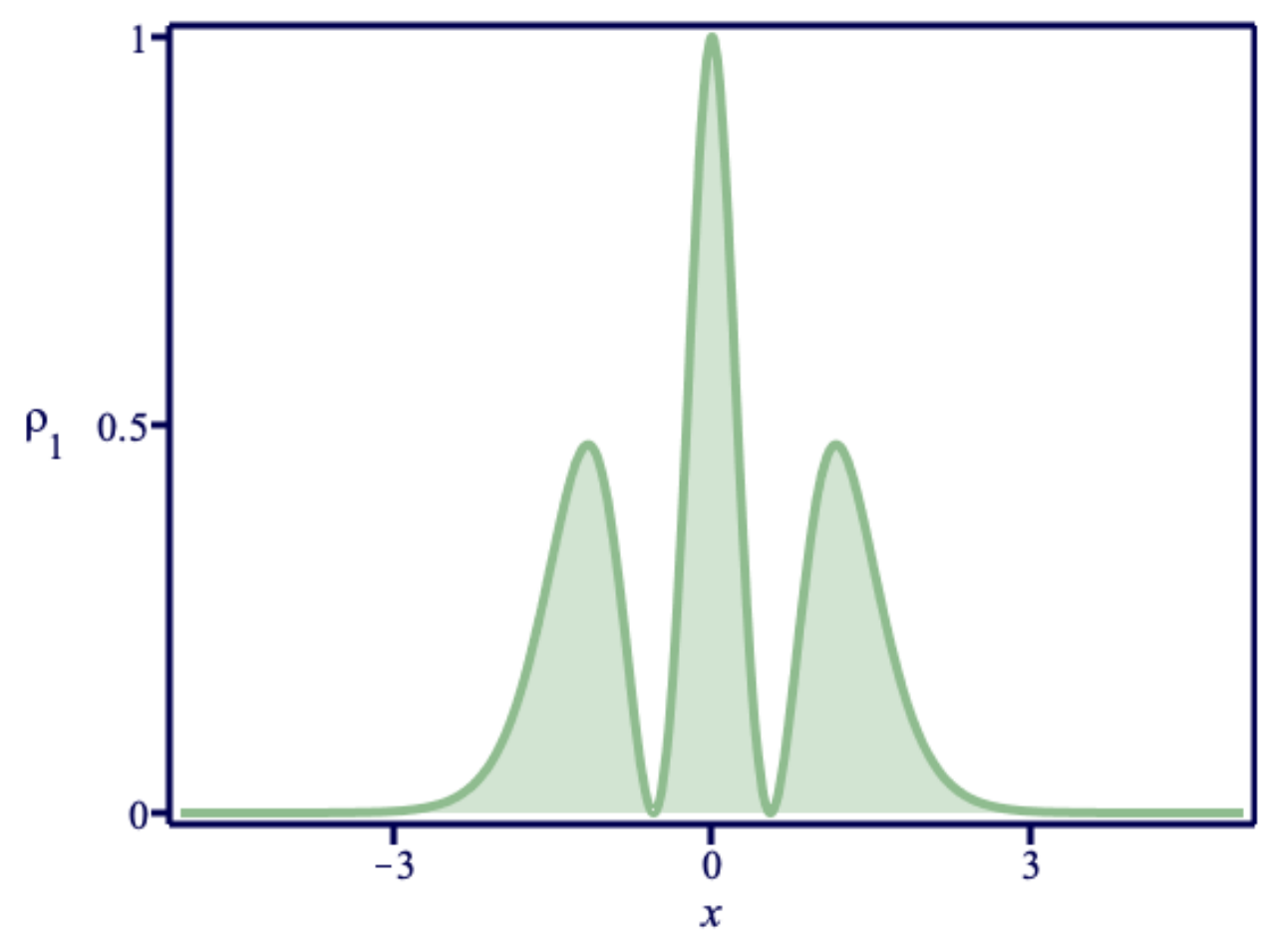}
\caption{Kinklike solution (top) and the corresponding energy density $\rho_1(x)$ (bottom) in the case $f(\chi)=1/\cos^2 (n\pi\chi)$, depicted for $n=1$ and $\alpha=0.5$ (left) and $1$ (right), respectively.}\label{fig3}}
\end{figure}
\begin{figure}[t!]
\centering{
\includegraphics[width=4.3cm]{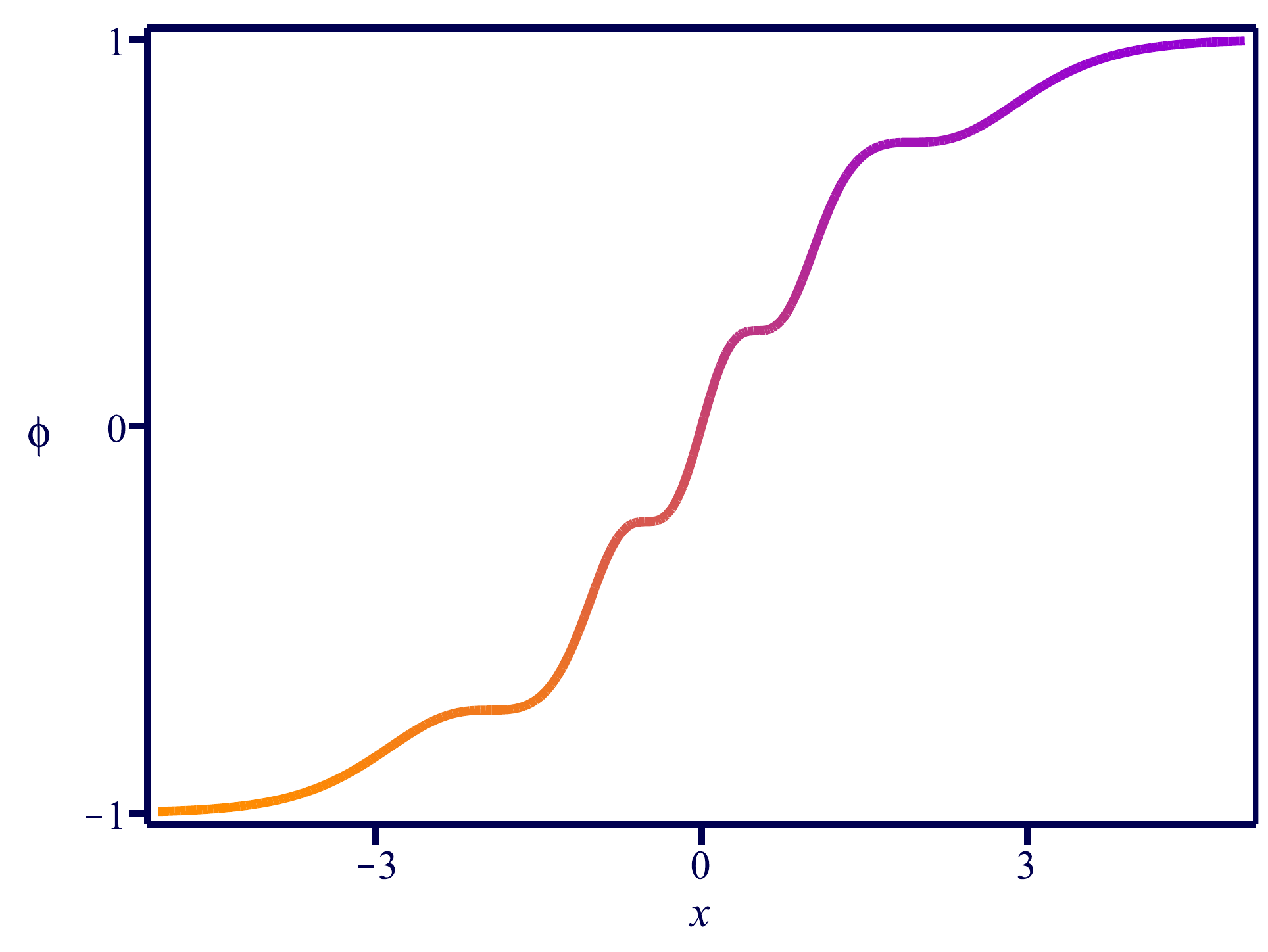}\includegraphics[width=4.3cm]{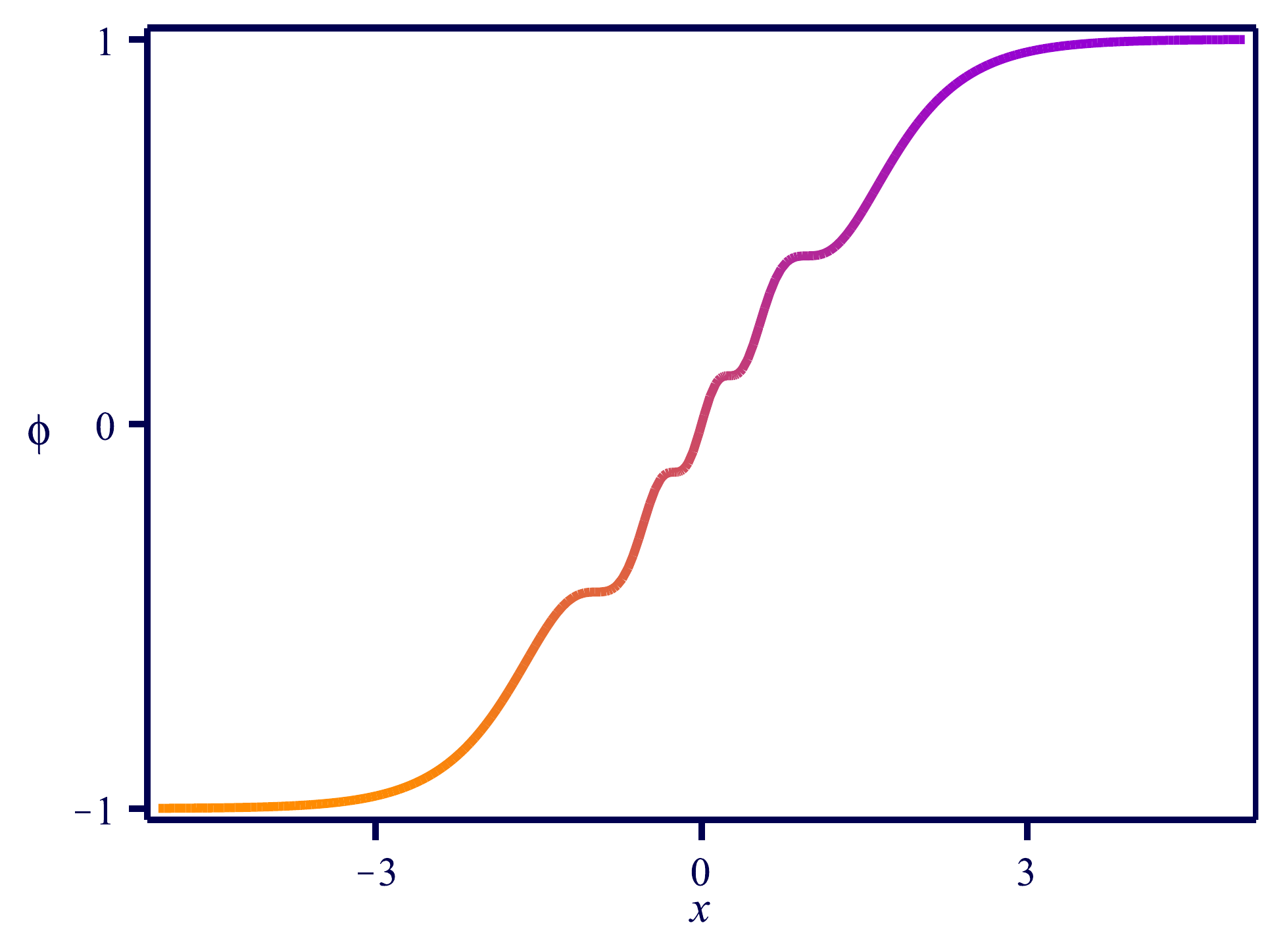}
\includegraphics[width=4.3cm]{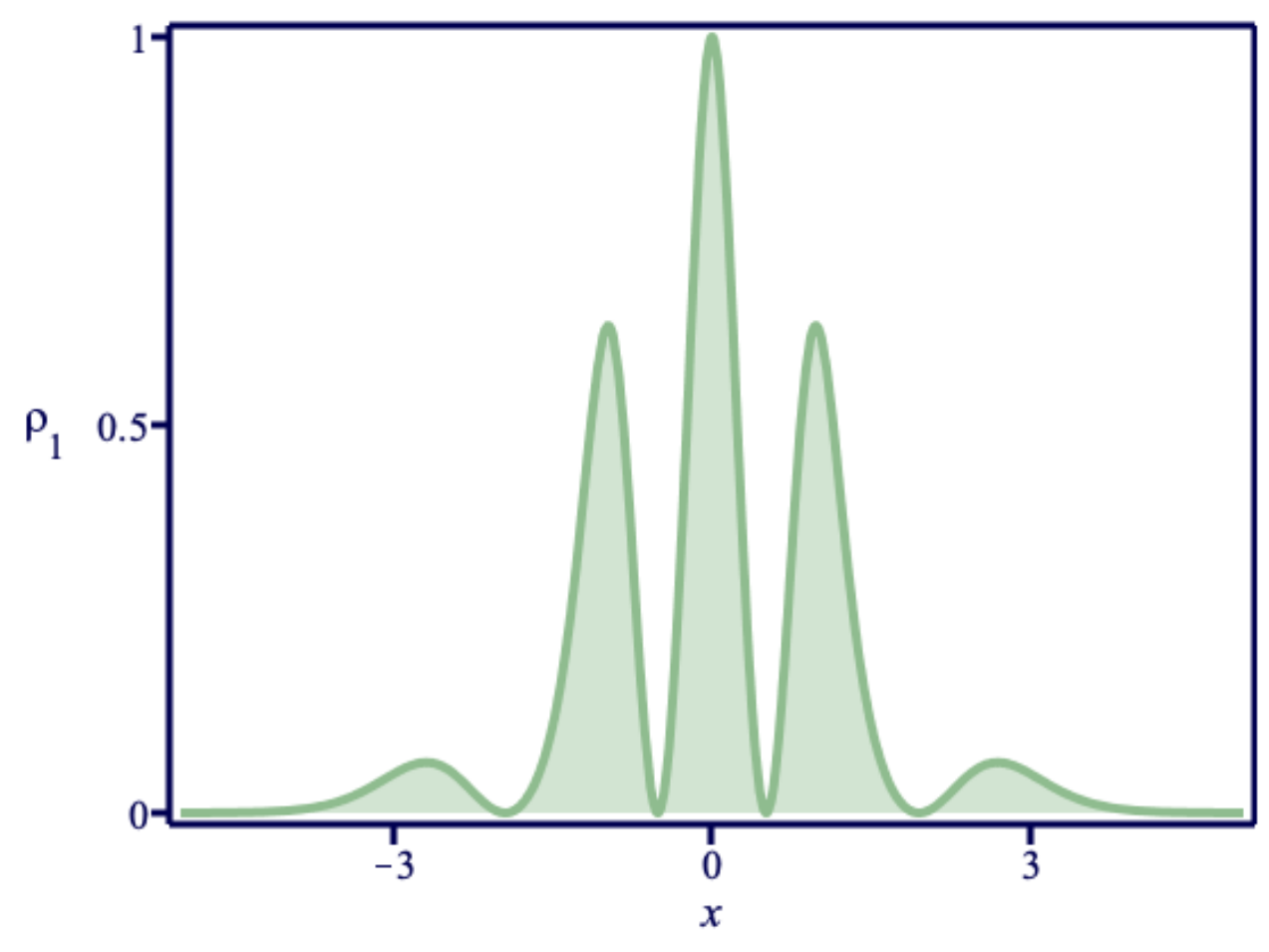}\includegraphics[width=4.3cm]{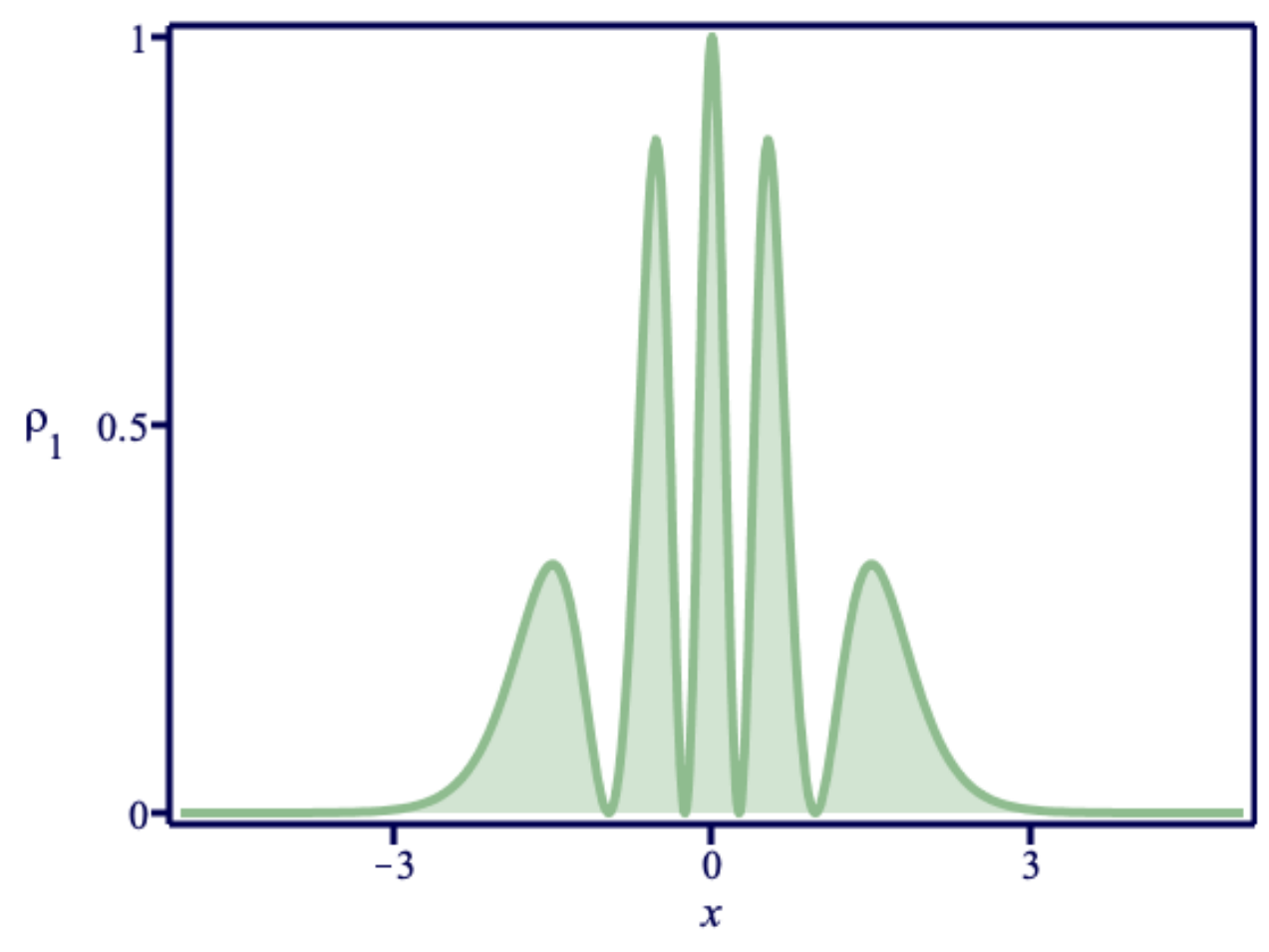}}
\caption{Kinklike solution (top) and the corresponding energy density $\rho_1(x)$ (bottom) in the case $f(\chi)=1/\cos^2 (n\pi\chi)$, depicted for $n=2$ and $\alpha=0.5$ (left) and $1$ (right), respectively.}\label{fig4}
\end{figure}

We investigate another model, described by the function $f(\chi)=1/\!\cos^2(n\pi\chi)$, $n\in\mathbb{N}$, and the function $W(\phi,\chi)$ given by Eq.~\eqref{wchi4}. In this case, the potential in Eq.~\eqref{potgen} becomes
	\begin{equation}
	V(\phi,\chi)=\frac12{\cos^2(n\pi\chi)}(1-\phi^2)^2+\frac{1}{2}\alpha^2(1-\chi^2)^2.
	\end{equation}
The minima are located at the values $\pm1$. As we can see from the first order procedure, the $\chi$ field and the energy density $\rho_2(x)$ contributes as in the previous case. However, the $\phi$ field obeys the new equation
	\be\label{fophimult}
	\frac{d\phi}{dx}= \cos^2(n\pi\tanh(\alpha x))\,(1-\phi^2).
	\ee
The solution is now a kinklike configuration with internal structure, which has the form
\begin{subequations}\label{multi}
  \begin{align}
\phi(x)&=\tanh\eta(x),\\
\eta(x)&=\frac{x}{2}+\frac{1}{4\alpha} \Big(\textrm{Ci}(\xi_n^+(x)) -\textrm{Ci}(\xi_n^-(x))\Big),
\end{align}
\end{subequations}
where $\textrm{Ci}(z)$ is the cosine integral function of $z$ argument, and
$\xi_n^\pm(x) =2n\pi(1\pm\tanh(\alpha x))$. {The function $\textrm{Ci}(z)$ is defined as
\be
\textrm{Ci}(z) = \gamma +\ln(z)+\int_0^z\,\frac{\cos(y)-1}{y}\,dy,
\ee 
where $\gamma=0.577$ is the Euler-Mascheroni constant. For small $z$, one gets 
\be 
\textrm{Ci}(z)=\gamma +\ln(z)-z^2/4+ \mathcal{O}(z^4).
\ee
On the other hand, for very large $z$, one has 
\be
\textrm{Ci}(z)=\sin(z)/z-\cos(z)/z^2 + \mathcal{O}(1/z^3).
\ee

Near the origin, one can show that the solution \eqref{multi} behaves as $\phi(x)\propto x$. The energy density $\rho_1$ in Eq.~\eqref{foeq} reads 
\be
\rho_1(x)=\cos^2(n\pi\tanh(\alpha x))\; \sech^4\eta(x),
\ee
One may integrate it to get $E_1 = 4/3$. We display the solution and the energy density $\rho_1(x)$ in Fig. \ref{fig3} for $n=1$ and $\alpha=0.5$ and $1$, and in Fig.~\ref{fig4} for $n=2$ and $\alpha=0.5$ and $1$. We note the presence of an even number of plateaux that inhabit the kink, which is controlled by $2n$.

If instead of the above $f(\chi)$ we take the new function $f(\chi)=1/\sin^2\big((n+1/2)\pi\chi\big)$, the model changes and although the solution is still given by \eqref{multi}, now we have to use $\xi_n^\pm(x) =(2n+1)\pi(1\pm\tanh(\alpha x))$. In this case, the 
configuration attains an odd number of plateaux, $2n+1$, with the profile of a two-kink around its center, behaving as $\phi(x)\propto \alpha ^2 x^3$ near the origin. In Fig. \ref{fig5} we display the kinklike configuration and the energy density $\rho_1(x)$ for $n=1$ and $2$, and for $\alpha=0.5$. 

		\begin{figure}[h!]
		\centering
		\includegraphics[width=4.2cm]{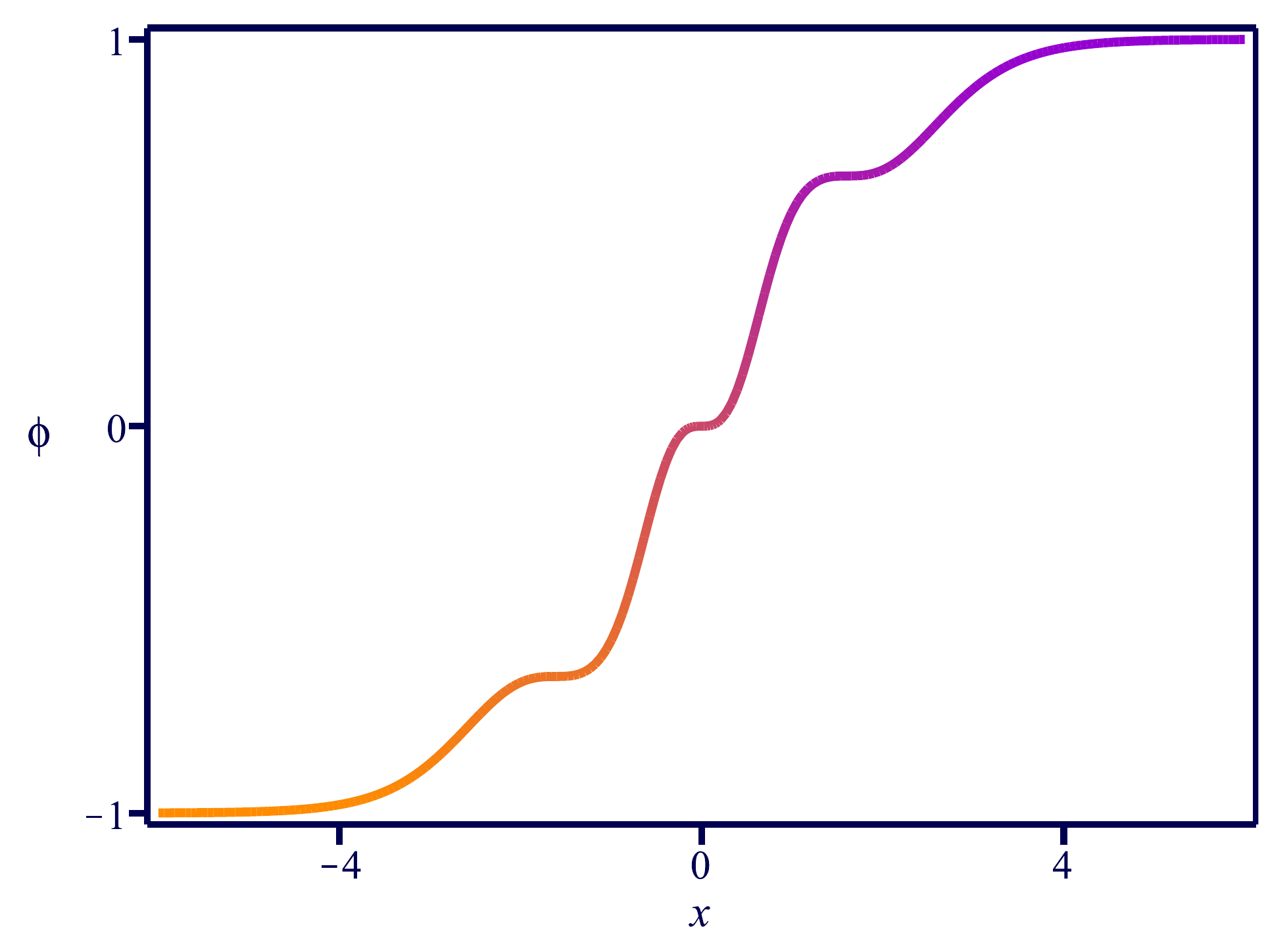}\includegraphics[width=4.2cm]{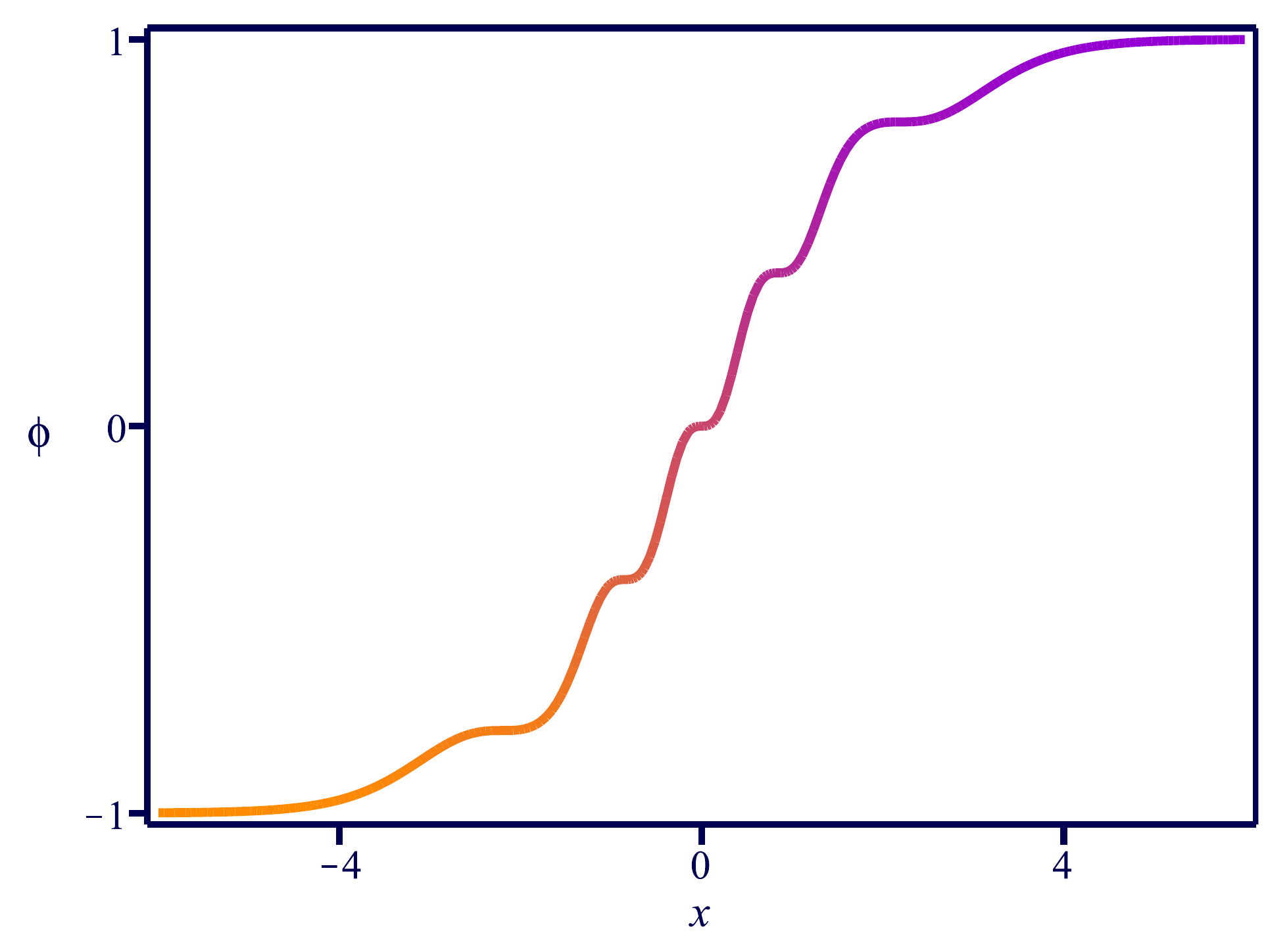}
		\includegraphics[width=4.2cm]{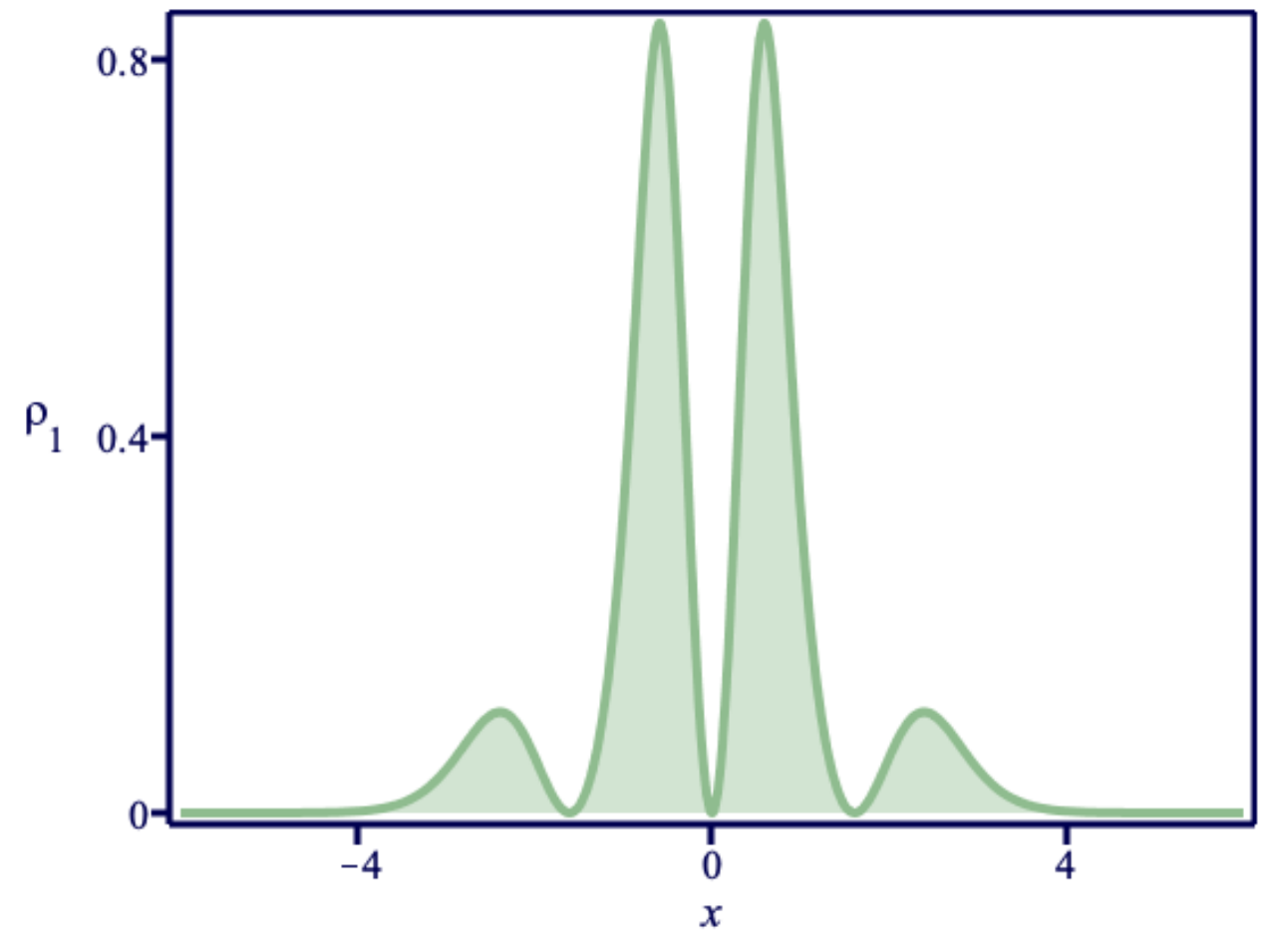}\includegraphics[width=4.2cm]{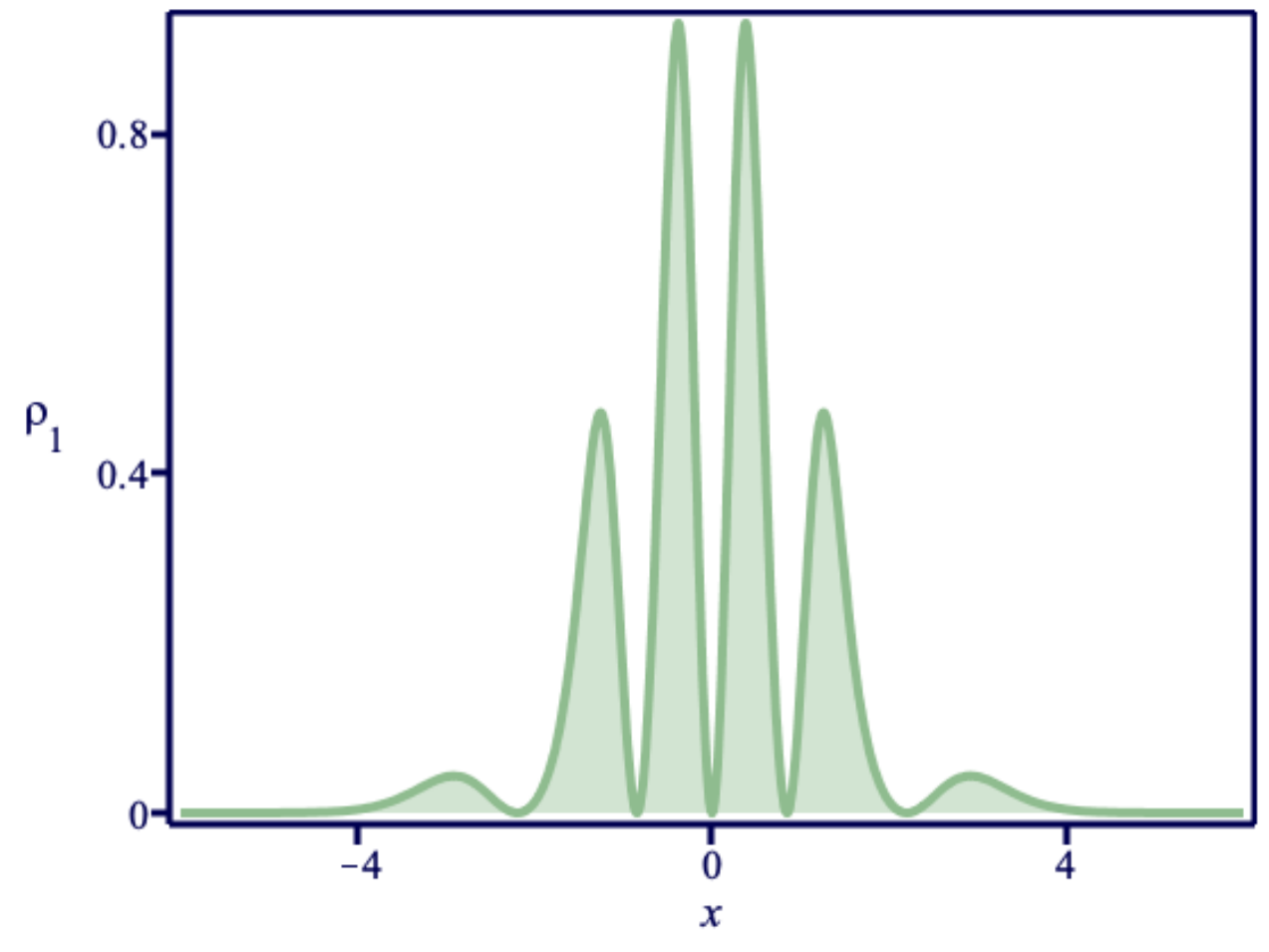}
		\caption{Kinklike solution (top) and the energy density $\rho_1(x)$ (bottom) in the case $f(\chi)=1/\sin^2\big((n+1/2)\pi\chi\big)$, depicted for $n=1$ (left) and $2$ (right), and for $\alpha=0.5$, respectively.}
		\label{fig5}
		\end{figure}

In summary, we have studied a model with $Z_2\times Z_2$ symmetry, with the extra $Z_2$ symmetry used to accommodate an additional field, which may give rise to kinks that modify the behavior of the other field. We have introduced an auxiliary function $W(\phi,\chi)$ of the fields and developed a first order framework that minimizes the energy of the system, leading to solutions that are stable against small fluctuations of the fields, and also, against rescaling of the spatial coordinate. With the help of the second field $\chi$ and the function $f(\chi)$, we have been able to show that the field $\phi$ may describe kinks with interesting internal behavior, similar to the effect found in \cite{KCM1} for the magnetization under the presence of a geometric constriction that leads to the two-kink profile, and in \cite{KCM2a} for the voltage in the longitudinal spin Seebeck effect, which also engenders the two-kink feature. We have examined another possibility, which gives rise to kinklike configurations, simulating the current that appears in the study of the quantum transport in molecular junctions, specially when one includes vibrational degrees of freedom \cite{O2,O3}.

We hope the above results motivate other investigations, in particular, on the behavior of the magnetization in the magnetic material examined in \cite{KCM1}, but now with an array of geometric constrictions, and also, on kink in the buckled graphene nanoribbon described in \cite{KCM4}, in the presence of geometric constrictions along the ribbon. These possibilities could give rise to configurations similar the ones investigated in this work. Another situation can be engineered to make the voltage in the longitudinal spin Seebeck effect to engender the novel kinklike profile, an issue of current interest to spintronics. The work fosters other studies, in particular the possibility to work in higher dimensions, focusing on planar and spatial structures such as skyrmions, vortices and magnetic monopoles.

\acknowledgements{This work is partially financed by Conselho Nacional de Desenvolvimento Cient\'\i fico e Tecnol\'ogico (CNPq grants 306614/2014-6, 404913/2018-0, 130923/2018-4, and 155551/2018-3) and by Paraiba State Research Foundation (FAPESQ-PB grant 0015/2019).}



\begin{thebibliography}{99}
\bibitem{B1} N. Manton and P. Sutcliffe, \textit{Topological solitons.} Cambridge University Press (2004).
\bibitem{B2}T. Vachaspati, \textit{Kinks and domain walls.} Cambridge University Press (2006).
\bibitem{B3}E.J. Weinberg, \textit{Classical solutions in quantum field theory.} Cambridge University Press (2012).
\bibitem{P}D. Finkelstein, J. Math. Phys. {\bf7}, 1218 (1966).
\bibitem{K0}D. Bazeia, L. Losano, and J. M. C. Malbouisson, Phys. Rev. D 66, 101701(R) (2002). 
\bibitem{K1} D. Bazeia, J. Menezes, and R. Menezes, Phys. Rev. Lett. {\bf91}, 241601 (2003).
\bibitem{K2}S. Dutta, D. A. Steer, and T. Vachaspati, Phys. Rev. Lett. {\bf101}, 121601 (2008).
\bibitem{K3}A. Alonso-Izquierdo, M. A. Gonzalez Leon, and J. Mateos Guilarte, Phys. Rev. Lett. {\bf101}, 131602 (2008).
\bibitem{K4}D. A. Takahashi and M. Nitta, Phys. Rev. Lett. {\bf110}, 131601 (2013).
\bibitem{K5}D. Matsunami, L. Pogosian, A. Saurabh, and T. Vachaspati,  Phys. Rev. Lett. {\bf122}, 201301 (2019).
\bibitem{KCM1}P.-O. Jubert, R. Allenspach, and A. Bischof, Phys. Rev. B {\bf69}, 220410(R) (2004).
\bibitem{KCM2}A. Vanhaverbeke, A. Bischof, and R. Allenspach, Phys. Rev. Lett. {\bf101}, 107202 (2008).
\bibitem{KCM2a}K. Uchida, H. Adachi, T. Ota1, H. Nakayama, S. Maekawa, and E. Saitoh, Appl. Phys. Lett. {\bf97}, 172505 (2010).
\bibitem{KCM3}F.J. Buijnsters, A. Fasolino, and M.I. Katsnelson, Phys. Rev. Lett. {\bf113}, 217202 (2014).
\bibitem{KCM4}R. D. Yamaletdinov, V. A. Slipko, and Y. V. Pershin, Phys. Rev. B {\bf96}, 094306 (2017).
\bibitem{O1}A. Mitra, I. Aleiner, and A. J. Millis, Phys. Rev. B {\bf69}, 245302 (2004).
\bibitem{O2}J. M. Thijssen and H. S. J. Van der Zant, Phys. Status Solidi B {\bf245}, 1455 (2008).
\bibitem{O3}M. Thoss and F. Evers, J. Chem. Phys. {\bf148}, 030901 (2018).
\bibitem{O4}P. Gehring, J. M. Thijssen, and H. S. J. van der Zant, Nature Review Physics {\bf1}, 381 (2019). 
\bibitem{O5}G. Erdemci-Tandogan, H. Orland, and R. Zandi, Phys. Rev. Lett. {\bf119}, 188102 (2017).
\bibitem{H} R. Hobart, Proc. Phys. Soc. Lond. {\bf82}, 201 (1963).
\bibitem{D} G.H. Derrick, J. Math. Phys. {\bf5}, 1252 (1964).
\bibitem{sky1}Y. Zhou and M. Ezawa, Nat. Commun. {\bf5}, 4652 (2014).
\bibitem{sky2}W. Jiang, P. Upadhyaya, W. Zhang, G. Yu, M. B. Jungfleisch, F. Y. Fradin, J. E. Pearson, Y. Tserkovnyak,
K. L. Wang, O. Heinonen, S. G. E. te Velthuis, and A. Hoffmann, Science {\bf349}, 283 (2015).
\bibitem{O6}C. Schinabeck, A. Erpenbeck, R. Hartle, and M. Thoss, Phys. Rev. B {\bf94}, 201407(R) (2016).
\bibitem{B}E.B. Bogomol'nyi, Sov. J. Nucl. Phys. {\bf24}, 449 (1976).
\bibitem{genkink} D. Bazeia, L. Losano, R. Menezes and J.C.R.E. Oliveira, Eur. Phys. J. C {\bf51}, 953 (2007).
\bibitem{genkink2} D. Bazeia, L. Losano and R. Menezes, Phys. Lett. B {\bf668}, 246 (2008).
\end{thebibliography}
\end{document}